\documentclass[twocolumn,reprint,floats,aps,amsmath,amssymb]{revtex4-1}
\usepackage[dvipsnames]{xcolor}
\usepackage{graphicx}% Include figure files
\usepackage{graphics}
\usepackage{float}
\usepackage[export]{adjustbox}
\usepackage{dblfloatfix}
\graphicspath{ {./images/} }
\usepackage{dcolumn}% Align table columns on decimal point
\usepackage{bm}% bold math
\everymath{\displaystyle}

\begin{document}

%\preprint{APS/123-QED}

\title{Nonlinear modulation of periodic waves in the cylindrical Gardner equation}
\author{G. Aslanova}
\email{Corresponding author: aslanova15@itu.edu.tr}
\author{S. Ahmetolan}
\email{ahmetola@itu.edu.tr}
 \author{A. Demirci}
 \email{demircial@itu.edu.tr}
\affiliation{
% Authors' institution and/or address\\
 %This line break forced with \textbackslash\textbackslash
 Department of Mathematics, Istanbul Technical University, Istanbul 34469, Turkey
}

\date{\today}% It is always \today, today,
             %  but any date may be explicitly specified

\begin{abstract}
The propagation of the dispersive shock waves (DSWs) is investigated in the cylindrical Gardner (cG) equation, which is obtained by employing  a similarity reduction to the two space one time (2+1) dimensional Gardner-Kadomtsev-Petviashvili (Gardner-KP) equation. We consider the step-like initial condition along a parabolic front. Then, the cG-Whitham modulation system, which is a description of DSW evolution in the cG equation, in terms of appropriate Riemann type variables is derived. Our study is supported by numerical simulations. The comparison is given between the direct numerical solution of the cG equation and the DSW solution obtained from the numerical solution of the Whitham system. According to this comparison, a good agreement is found between the solutions. 
%\begin{description}
%May be entered using the \verb+\pacs{#1}+ command.
%\item[Structure]
%You may use the \texttt{description} environment to structure your abstract;
%use the optional argument of the \verb+\item+ command to give the category of each item. 
%\end{description}
\end{abstract}

%\pacs{Valid PACS appear here}% PACS, the Physics and Astronomy
                             % Classification Scheme.
%\keywords{Suggested keywords}%Use showkeys class option if keyword
                              %display desired
\maketitle

%\tableofcontents

\section{\label{sec:level1}INTRODUCTION}

The Gardner equation
\begin{equation}
\label{eq1}
   u_{t}+6uu_{x}\pm6 u^2u_{x}+u_{xxx}=0
\end{equation}
is a well-known model which describes internal solitary waves in shallow water and it is called the focusing or defocusing Gardner equation depending on the plus or minus sign of the cubic nonlinear term, respectively.
This equation is firstly derived to obtain the infinite set of local conservation laws of the Korteweg-de Vries (KdV) equation \cite{Gardner}. It was later accepted as a universal model for describing nonlinear wave propagation in layered fluids. This equation has importance in modeling large amplitude inner waves in the ocean (see \cite{Grimshaw2002, Helfrich2006, Apel2007}). It also describes a variety of wave phenomena in solid-state and plasma physics \cite{Watanabe1984, Ruderman2008}, dynamics of Bose-Einstein condensate (BEC) \cite{BEC-K} and quantum field theory \cite{Demler2011}. In addition, it was shown that the Benjamin-Bona-Mahoney (BBM) equation with power law nonlinearity can be transformed to the combined KdV-mKdV equation, which is also known as the Gardner equation \cite{Johnpillai2013}. \par
The Gardner equation is extended to the Gardner-KP equation by using the sense of the Kadomtsev and Petviashvili \cite{KP1970}, who relaxed the restriction that the waves be absolutely one-dimensional. In this study, we are interested in a defocusing-type Gardner-KP equation in the following form:
\begin{equation}
\label{eq2}
    (u_{t}+6uu_{x}-6u^2u_{x}+\epsilon^2u_{xxx})_{x}+\lambda u_{yy}=0,	
\end{equation}	
where, $0<\epsilon<<1$ and $\lambda=\pm1$ are constants. The Gardner-KP equation describes strong nonlinear internal waves on ocean shelf in two dimensional case. This equation has two nonlinear terms in the quadratic and cubic forms and the dispersive term is of third order.  \par

In this study, we consider the formation and the propagation of a dispersive shock waves in the cG equation (see Eq. \eqref{eqB11}), which is obtained by using a similarity reduction to the Gardner-KP equation \eqref{eq2}.
Dispersive shock waves (DSWs), also termed undular bores in fluid mechanics, are slowly modulated non-stationary wavetrains that develop spontaneously in weakly dispersive nonlinear media. In this waveform, the nonlinearity induces front steepening and thus induces the tendency to develop an unphysical hydrodynamic singularities, named gradient catastrophe. A weak dispersion takes the second place until steep gradients are eventually formed. At this stage dispersion becomes effective. The result is the expanding front characterized by oscillations. These oscillations spread in a characteristic fan in the space-time plane and the borders of this fan represent the leading and the trailing edge of the DSW, where the amplitude of the oscillations are largest and vanishingly small, respectively. These two edges propagate with different speeds.
\par
The investigation of DSWs has a long history that begins with Whitham's pioneering invention of modulation theory \cite{Whitham1965, Whitham1974} and continues with the construction of the DSW solution for the KdV equation, which physically describes an undular bore by Gurevich and Pitaevskii \cite{Gurevich}. It was then verified numerically by Fornberg and Whitham \cite{Fornberg1978}. The modulation theory for the Gardner equation was developed in \cite{Kam12}, where the complete classification of the solutions for the Riemann type problem was constructed. \par
In this study, a multiple-scale method \cite{Luke} is used to investigate DSWs in the cylindrical Gardner equation as an alternative to the method applied by Whitham. By using this method, a system of quasilinear modulation equations describing slow evolution of parameters in the periodic travelling wave solution, such as amplitude, wave number and mean height are obtained. These equations are called Whitham modulation equations. For hyperbolic systems, a DSW solution occurs as a stable wavetrains and for elliptic systems it corresponds to an unstable wavetrains. When the connection between DSW solutions and hyperbolic modulation equations was realised, the DSW solutions of other integrable nonlinear, dispersive wave equations, such as the nonlinear Schrödinger (NLS) equation \cite{Forest1986, Pavlov1987}, the modified KdV (mKdV) equation  \cite{Driscoll1976}, the KdV-Burgers equation \cite{Gurevich1987}, the Benjamin-Ono equation \cite{Matsuno} and the Gardner equation \cite{Kam12} were found.\par
The key feature to find the DSW solution from modulation equations is the ability to set them in the appropriate Riemann variable form, which is guaranteed if the underlying equation is integrable. However, most equations governing DSWs in physical applications are not integrable. Then, based on Whitham’s and Gurevich and Pitaevskii’s research, El proposed the framework of a DSW fitting method which enables the analysis of DSWs governed by non-integrable equations \cite{El2005, El2003}. This method was used to find the leading (solitary wave) and trailing (linear wave) edges. In \cite{Kam_non}, Kamchatnov also investigated DSWs in non-integrable equations.  \par
All of these studies were restricted to (1+1) dimensional PDEs and much less information had been known about DSWs in multidimensional PDEs until recently. In the last decade, DSWs in two space one time (2+1) dimensional systems have been subject of few studies. In \cite{Demirci2016}, by using a similarity variable, (1+1) dimensional cylindrical reductions of the Kadomtsev-Petviashvili and two dimensional Benjamin-Ono equations and their associated DSW solutions were investigated. We note that the method used in \cite{Demirci2016} works only under the special choice of parabolic front. Later, a generalization of Whitham theory for DSWs in the KP-type equations with a general class of initial conditions was developed \cite{Ablowitz18}. The main result of this work was the derivation of the system of (2+1) dimensional hydrodynamic-type equations, which describes the slow modulations of the periodic solutions of the corresponding KP-type equation. The method presented in \cite{Ablowitz18} can be applied to DSW investigation in integrable and non-integrable (2+1) dimensional PDEs. Actually, the Gardner-KP equation \eqref{eq2} belongs to the equation class studied in \cite{Ablowitz18}. However, quantitative results obtained from our present study about DSWs in the Gardner-KP equation will be very important to verify the theoretical results of the study in \cite{Ablowitz18}.
\par
In this study, we employ a parabolic similarity reduction to the integrable (2+1) dimensional Gardner-KP and then this equation reduces to (1+1) dimensional equation called the cylindrical Gardner (cG) equation. After that, we apply the Whitham modulation theory to the cG equation. We obtain two secularity (compatibility) conditions associated with the Whitham modulation theory for the cG equation by using a perturbation method. Then, these two secularity conditions with consistency requirement (conservation of waves) are used to derive a system of three quasilinear first order partial differential equations. This system is called Whitham modulation system and describes the modulations of the travelling wave solutions of the cG equation. By introducing appropriate Riemann type variables, the corresponding modulation equations are transformed into simpler form. This simpler form is important to understand the dispersive shock wave phenomena in the cG equation. Next, we obtain the direct numerical solution of the cG equation and compare this solution with the DSW solution. The analytical results from the modulation theory are shown to be in good agreement with direct numerical solution of the cG equation.\par
The paper is organized as follows. In Section \ref{sec:level2}, we consider the Gardner-KP equation. A special ansatz is used for the Eq. (\ref{eq2}), then the (2+1) dimensional equation reduces to the cylindrical Gardner equation. In Section \ref{sec:level3}, we derive the modulation equations in terms of three Riemann type variables. The modulation equations have the form of a hyperbolic system of PDEs for the slowly varying parameters of the travelling wave solution. The Section \ref{sec:level4} deals with the direct numerical solution of the cG equation. In Section \ref{sec:level5}, the formation of DSW in the cG equation is considered, based on the results obtained in Section \ref{sec:level3}. Also, both the direct numerical solution of the cG equation and the DSW solution obtained from the numerical solution of the corresponding Whitham system is compared. In addition, to observe the effect of the cylindrical term, the same examinations were carried out for the Gardner equation, inspired by the study \cite{Kam12}. Finally, all results are summarized in Section \ref{sec:level6}. 

%**************************************************************
\section{\label{sec:level2}THE CYLINDRICAL GARDNER EQUATION}

In this section, the cG equation is presented by reducing the (2+1) dimensional Gardner-KP equation (\ref{eq2}) with a parabolic similarity reduction \cite{Demirci2016}.
We are interested in a class of initial conditions for the Gardner-KP equation describing almost step-like initial data as follows: 
\begin{equation}
\label{eqB2}
    u(x,y,0)\!\!=\!\!\frac{1}{2}
    [(f^{-}\!\!+\!f^{+})\!\!+\!(f^{+}\!\!-f^{-})
    \tanh(A(x\!\!+\!\frac{1}{2}\phi(y,0)))
    ]
\end{equation}
where $f^{-}$, $f^{+}$ and $A$ are real constants. $\phi(y,t)$ describes the front shape of the solution of Eq. (\ref{eq2}). In this study, we choose a parabolic front $\phi(y,0)=\tilde{c}y^2$ where $\tilde{c}$ is a real constant.\par
Ablowitz et al. have used a reduction method for the above initial data type to describe the dispersive shock waves in the Kadomtsev-Petviashvili and two dimensional Benjamin-Ono equation \cite{Demirci2016}. The method used in \cite{Demirci2016} works under the special choice of a parabolic front or a planary front. However, the only way to obtain the cylindrical equation from the reduction of (2+1) equation is to take the parabolic front. \par

We use the following ansatz:
\begin{equation}
\label{eqB3}
    u=f\Big(x+\frac{\phi(y,t)}{2},t,y\Big)  	
\end{equation}
for the Gardner-KP equation (\ref{eq2}), where the front is then described by $x+\phi(y,t)/2=constant$. When we substitute the ansatz (\ref{eqB3}) into Eq. (\ref{eq2}) we obtain
\begin{equation}\label{eqB4}
\begin{array}{lcl}
   &\Big(\frac{1}{2}\phi_{t}f_{\eta}+f_{t}+6ff_{\eta}-6f^2 f_{\eta}+\epsilon^2 f_{\eta\eta\eta}\Big)_{\eta}\\
   +&\lambda\Big(\frac{1}{4}(\phi_{y})^2f_{\eta\eta}+\frac{1}{2}\phi_{yy}f_{\eta}+\phi_{y}f_{\eta y}+f_{yy}\Big)=0,	
\end{array}
\end{equation}
where $\eta=x+\phi(y,t)/2$. Also, we assume that $u$ satisfies the following boundary conditions at the infinities with $f^{-}>f^{+}\geq 0$ for non-increasing type initial conditions:
\begin{equation}
\label{eqB5}
   u\to R(t)f^- \text{ as } \eta\to-\infty \text{ and } u\to R(t)f^+ \text{ as } \eta\to\infty. 
\end{equation}
The function $R(t)$ will be determined at the end of this section with the initial condition $R(0)=1$.\par
Assuming that $\phi_{yy}$ is independent of $y,$ due to the assumption of the parabolic front, the system of equations in the following form is obtained:
\begin{eqnarray}
\label{eqB6}
\phi_{t}+\frac{\lambda}{2}(\phi_{y})^2 &=&0,\\
f_{t}+6f f_{\eta}-6f^{ 2}f_{\eta}+\frac{\lambda}{2}\phi_{yy}f+\epsilon^2 f_{\eta\eta\eta}&=&0.
\end{eqnarray}
We call Eq. (7) as the front shape equation which describes the evolution of the curvature of the parabolic front. However, Eq. (8) characterizes dispersive shock wave propagation of the wave front. Eq. (7) can be transformed to the Hopf equation by using the transformation $v=\phi_{y}$:
\begin{equation}
\label{eqB8}
v_t+\lambda v v_y=0.
\end{equation}
The solution of Eq.(9) with the initial condition $v(y,0)=2\tilde{c}y$ is
\begin{equation}
\label{eqB9}
v(y,t)=\frac{2\tilde{c}y}{1+2\tilde{c}\lambda t}.
\end{equation}
Thus the front shape function $\phi(y,t)$ is obtained  as
\begin{equation}
\label{eqB10}
\phi(y,t)=\frac{\tilde{c}y^2}{1+2\tilde{c}\lambda t}.
\end{equation}
The substitution of (\ref{eqB10}) into Eq. (8) gives the following cG equation; 
\begin{equation}
\label{eqB11}
f_{t}+6f f_{\eta}-6f^{2}f_{\eta}+\frac{\lambda\tilde{c}}{1+2\tilde{c}\lambda t}f+\epsilon^2 f_{\eta\eta\eta}=0.
\end{equation}
Denoting $t_0=1/\lambda\tilde{c},$ the term $\lambda\tilde{c}/({1+2\tilde{c}\lambda t})$ transforms to $1/(2t+t_0).$ We will consider $\lambda=1$. The other sign can be obtained by changing the $\tilde{c}$ to $-\tilde{c}$, i.e. changing the direction of the parabolic front.
\par
Note that, there is another possibility of the choice for the initial front. When the front is chosen planary as $\phi(y,0)=\tilde{c}y$, then $\phi_{yy}$ is independent of $y$ and the form of Eq. (8) becomes the classical Gardner equation. DSWs in the Gardner equation was studied in \cite{Kam12}.
\par 
We construct the DSW solution of the cG equation \eqref{eqB11} with the step type initial condition in the following form:
\begin{equation}
\label{eqB12}
f(\eta,0) = \left\{ \begin{array}{ll}
f^{-}, & \textrm{$\eta<0;$}\\
f^{+}, & \textrm{$\eta>0.$}
\end{array} \right.
\end{equation}
\par
The structure of the waves modeled by Gardner and cylindrical Gardner equations depends on the values of the initial step parameters $f^{-}, f^{+}$. We require that $f^->f^+\geq0 $ in order for the generation of a DSWs in cG equation. In this study, we examine the equation (\ref{eqB11}) with considering required condition for DSW formation in the Gardner equation \cite{Kam12}. \par
Now to find the function $R(t)$ in the boundary conditions \eqref{eqB5}, first we neglect $\eta$ dependent terms in Eq. \eqref{eqB11} and get an ordinary differential equation (ODE). The solution of this ODE with the initial condition $R(0)=1$ determines the function $R(t)$ in the boundary conditions (\ref{eqB5}) as
\begin{equation}
\label{eqB13}
R(t)=\frac{1}{\sqrt{1+2\tilde{c}t}}.
\end{equation}
	
%*****************************************************************

\section{\label{sec:level3}DERIVATION OF THE MODULATION EQUATIONS}
The key to obtaining DSW solutions is the usage of the Whitham modulation theory \cite{Whitham1965, Whitham1974}. We construct a DSW solution of the cG equation by using the method of multiple-scale for analysing slowly varying, nonlinear dispersive waves which is different from the original usage of the Whitham theory. The multiple-scale method used in this study was initially introduced by Luke \cite{Luke}. By using this method, the Whitham modulation equations describing a system of PDEs for the slowly varying parameters of a periodic travelling wave solution, such as amplitude, wavenumber and mean height are constructed. These equations are important to understand the dispersive shock wave phenomena in the cG equation.\par
A DSW consists of two edges, the trailing edge and leading edge with a modulated dispersive wavetrain between these edges. These two edges move with different speeds. Also, the trailing edge corresponds to the small amplitude sinusoidal wave train, while the leading edge corresponds to the large amplitude solitary waves. \par
According to the modulation theory, wave parameters change slowly over fast oscillations within the DSW. This requirement can be formalized by introducing a rapidly varying phase variable, where
\begin{equation}
\label{eqC3}
\theta_{\eta}=\frac{k}{\epsilon},\quad \theta_{t}=-\frac{\omega}{\epsilon}=-\frac{kV}{\epsilon}.
\end{equation}
Here $\eta$, $t$ are slow space-time variables and $k(\eta,t)$, $\omega(\eta,t)$ and $V(\eta,t)$ are the wave number, frequency and the phase velocity, respectively. We assume that $0<\epsilon<<1$.\par

Since $(\theta_{\eta})_{t}=(\theta_{t})_{\eta}$ we obtain the  compatibility condition (conservation of waves) as follows:
\begin{equation}
\label{eqC4}
k_{t}+(kV)_{\eta}=0.
\end{equation}
\par
By the following relations,
\begin{equation}
\label{eqC5}
\frac{\partial}{\partial\eta}\to \frac{k}{\epsilon} \frac{\partial}{\partial\theta}+\frac{\partial}{\partial\eta},\quad \frac{\partial}{\partial t}\to -\frac{\omega}{\epsilon} \frac{\partial}{\partial\theta}+\frac{\partial}{\partial t}
\end{equation}
Eq. (\ref{eqB11}) is transformed to 
\begin{equation} 
\label{eqC6}
\begin{split}
&\Big(\!-\frac{\omega}{\epsilon}\frac{\partial}{\partial\theta}\!+\!\frac{\partial}{\partial t}\Big)f\!+\!6f\Big(\frac{k}{\epsilon} \frac{\partial}{\partial\theta}\!+\!\frac{\partial}{\partial\eta}\Big)f\!+\!\frac{\lambda\tilde{c}}{1\!+\!2\tilde{c}\lambda t}f\\
&-6f^2\Big(\frac{k}{\epsilon}\frac{\partial}{\partial\theta}+\frac{\partial}{\partial\eta}\Big)f+\epsilon^2\Big(\frac{k}{\epsilon}\frac{\partial}{\partial\theta}+\frac{\partial}{\partial\eta}\Big)^3 f=0.
\end{split}
\end{equation}
Grouping the terms in like powers of $\epsilon$, we rewrite Eq. (\ref{eqC6}) as seen below:
\begin{equation} 
\label{eqC7}
\begin{split}
&\frac{1}{\epsilon}\Big(\!-\omega f_\theta\!+\!6kf f_\theta\!-6k f^2 f_\theta\!+\!k^3f_{\theta\theta\theta}\Big)\\
  &+\!\Big(f_t\!+\!6ff_\eta\!-6f^2f_\eta\!+3k^2f_{\eta\theta\theta}\!+\!3kk_\eta f_{\theta\theta}\!+\!\frac{\lambda\tilde{c}}{1\!+\!2\tilde{c}\lambda t}f\Big)\\
  &+\!\epsilon \big( 3kf_{\theta\eta\eta}\!+\!3k_\eta f_{\eta\theta}\!+\!k_{\eta\eta}f_\theta \big)\!+\!\epsilon^2 f_{\eta\eta\eta}=0.
\end{split}
\end{equation}
When we expand the function $f$ in powers of $\epsilon$ as 
\begin{equation}
\label{eqC2}
f(\theta,\eta,t)=f_0(\theta,\eta,t)+\epsilon f_1(\theta,\eta,t)+...,
\end{equation}
the leading and the next order perturbation equations are obtained as,
\begin{equation}
\label{eqC8}
O\Big(\frac{1}{\epsilon}\Big): -\omega f_{0,\theta}+6kf_0 f_{0,\theta}-6k f_0^2 f_{0,\theta}+k^3f_{0,{\theta\theta\theta}}=0,
\end{equation}
\begin{equation} \label{eqC9}
\begin{split}
O(1): &\!-\omega f_{1,\theta}\!+\!6k(f_0 f_1)_\theta\!-\!6k f_0^2 f_{1,\theta}\!-\!12kf_0f_1f_{0,\theta}\\
&\!+\!k^3f_{1,{\theta\theta\theta}}\!=U,
\end{split}
\end{equation}
where
\begin{equation} 
\label{eqC10}
\begin{split}
U= &-\Big(f_{0,t}+6f_0 f_{0,\eta}-6f_0^2 f_{0,\eta}+3k^2 f_{0,\eta\theta\theta}\\
&+3kk_{\eta} f_{0,\theta\theta}+\frac{f_0}{2t+t_0}\Big).
\end{split}
\end{equation}
We can proceed to higher order terms, but doing so is outside the scope of this paper.
\par 
In order to solve the leading order problem, the travelling wave solution of the defocusing Gardner equation is examined due to the similarity in the structure of this equation with Eq. (\ref{eqC8}). We consider the travelling wave ansatz $f=f(\xi),\quad \xi=x-Vt$ in Eq. (\ref{eq1}) with minus sign by taking $f$ instead of $u$ and integrate this equation twice with respect to $\xi$ to obtain
\begin{equation} \label{eqC10y}
f_\xi^2=f^4-2f^3+Vf^2+Af+B,
\end{equation}
where $A$ and $B$ are the constants of integration. The solution of this equation can be expressed in terms of the Jacobian elliptic functions $cn$ and $sn$. A cnoidal wave solution of the Gardner equation is stable if all roots of the right hand-side polynomial of Eq. \eqref{eqC10y} are all real, and unstable if two roots are real, two are complex. Supposing that all real roots of the corresponding right-hand side polynomial  $a_1, a_2, a_3, a_4,$ are ordered as  
\begin{equation}\label{eqC11y}
a_1<a_2<a_3<a_4,
\end{equation}
then a travelling wave solution exists for $a_2<f<a_3.$ In this case, the right hand side of Eq.  (\ref{eqC10y}) is written as 
\begin{equation} \label{eqC12y}
f^4\!-\!2f^3\!+\!Vf^2\!+\!Af\!+B\!=\!(f\!-\!a_1)(f\!-\!a_2)(a_3\!-\!f)(a_4\!-\!f),
\end{equation}
where
\begin{equation}\label{eqC13y}
%\begin{array}{lcl}
\begin{aligned}
a_1+a_2+a_3+a_4&=2\\
a_1a_2+a_1a_3+a_1a_4+a_2a_3+a_2a_4+a_3a_4&=V\\
-a_1a_2a_3-a_1a_2a_4-a_1a_3a_4-a_2a_3a_4&=A\\
a_1a_2a_3a_4&=B.
\end{aligned}
%\end{array}
\end{equation}
Therefore, three of $a_j$'s are independent. In a modulated wave, which we are interested in, they are slowly varying functions of space coordinate $\eta$ and time $t$, $a_i=a_i(\eta, t)$. Their evolution is governed by the Whitham modulation equations, which will describe dispersive shock wave formation of the cG equation.\par
Provided that $a_2<f<a_3$, Eq.(\ref{eqC10y}) can be rewritten formally as:
\begin{equation}\label{eqC14y}
    \frac{df}{\sqrt{(f-a_1)(f-a_2)(a_3-f)(a_4-f)}}=d\xi.
\end{equation}
If we integrate Eq. (\ref{eqC14y}), the solution of Eq. \eqref{eqC10y} which is also the solution of the leading order problem \eqref{eqC8},  in terms of Jacobi elliptic functions is obtained as
\begin{equation}
\label{eqC11}
f_0=a_2+\frac{(a_3-a_2)cn^2(2(\theta-\theta_0)K,m)}{1-\frac{a_3-a_2}{a_4-a_2}sn^2(2(\theta-\theta_0)K,m)}.
\end{equation}
Here $K=K(m)$ is the complete elliptic integral of the first kind and $m$ is the modulus of the elliptic function $cn$, where  \par
\begin{equation}
\label{eqC13}
m^2=\frac{(a_3-a_2)(a_4-a_1)}{(a_3-a_1)(a_4-a_2)}.
\end{equation}
Note that, there is a free constant $\theta_0$ in Eq. \eqref{eqC11}. It is possible to find $\theta_0$ by constructing Whitham equations to higher order in much the same way as one can develop for higher order KdV or nonlinear Schrödinger type equations in physical applications. Such higher order analysis is outside the scope of this paper. We determine the approximated value of $\theta_0$ by comparison with direct numerical solutions.
\par
As mentioned before our aim is to obtain the three modulation equations for the three independent parameters $a_2, a_3, a_4$ of the solution \eqref{eqC11}. $k$, $m$ and $V$ will be expressed in terms of these independent variables. One of these modulation equation is Eq.(\ref{eqC4}), which is the conservation of waves. To obtain the other two equations, the problem $O(1)$ given in Eq. (\ref{eqC9}) should be examined. If the leading order solution \eqref{eqC11} is used in Eq. (\ref{eqC9}), secular terms, arbitrarily large growing terms with respect to $\theta$ are occured. To eliminate these terms, we enforce the periodicity of $f_0$ in $\theta$ and obtain the secularity conditions as
\begin{equation}
\label{eqC16}
\int_0^1 Ud\theta=0 \hspace{0.3cm}\textrm{ and } \int_0^1 f_0Ud\theta=0.
\end{equation}
Replacing $U$ given in (\ref{eqC10}) into the Eqs. (\ref{eqC16}), we obtain 
\begin{equation}
\label{eqC17}
\frac{\partial}{\partial t}\int_0^1 f_0d\theta+\frac{\partial}{\partial \eta}\int_0^1 (3f_0^2-2f_0^3)d\theta+\frac{1}{2t+t_0}\int_0^1 f_0d\theta=0
\end{equation}
and
\begin{equation} 
\label{eqC18}
\begin{split}
&\frac{\partial}{\partial t}\int_0^1 f_0^2d\theta+\frac{\partial}{\partial \eta}\int_0^1 (4f_0^3-3f_0^4-3k^2f_{0,\theta}^2)d\theta\\
&+\frac{2}{2t+t_0}\int_0^1 f_0^2d\theta=0.
\end{split}
\end{equation}

Equations (\ref{eqC4}), (\ref{eqC17}) and (\ref{eqC18}) are the required modulation equations. If we calculate the functions $f_0^2, f_0^3, f_0^4, f_{0,\theta}^2$ and their integrals by using the properties of elliptic functions \cite{Handbook}, we can obtain modulation equations in terms of $a_2, a_3, a_4$ variables ($a_1$ is eliminated with the help of the first equation in (\ref{eqC13y})).
\par 
Note that, we need to use the derivative formulas (\ref{eqA5})-(\ref{eqA8}) of the elliptic integrals in the first, second and third types (see Appendix) to obtain the system of modulation equations which is a first order quasilinear PDE system with the following form 
\begin{equation}
\label{eqC19}
\mathbf{u}_t+A(\mathbf{u})\mathbf{u}_\eta+B(\mathbf{u})\frac{1}{2t+t_0}=0,
\end{equation}
where $\mathbf{u}(\eta,t)=(a_2,a_3,a_4),$ $A(\mathbf{u})$ is a $3\times3$ matrix and $B(\mathbf{u})$ is a $3\times1$ vector.

The modulation equations can be simplified by replacing the variables $a_2,a_3,a_4$ with $r_1,r_2,r_3$, $(r_1\leq r_2\leq r_3)$, where $r_i$,\quad$i=1,2,3$ are the Riemann variables. For the cG equation we can take Riemann variables as given below \cite{Kam12}:
\begin{equation}\label{eqC20}
\begin{aligned}
r_1=\frac{1}{4}(a_1+a_2)(a_3+a_4),
\\
r_2=\frac{1}{4}(a_1+a_3)(a_2+a_4),
\\
r_3=\frac{1}{4}(a_2+a_3)(a_1+a_4).
\end{aligned}
\end{equation}
Thus, the Whitham modulation system in Eq. (\ref{eqC19}) can be transformed to the simpler form:
\begin{equation}
\label{eqC21}
\frac{\partial r_i}{\partial t}+v_i(r_1,r_2,r_3)\frac{\partial r_i}{\partial \eta}+\frac{h_i(r_1,r_2,r_3)}{2t+t_0}=0,\quad i=1,2,3,
\end{equation}
where,
\begin{equation}\label{eqC22}
\begin{aligned}
v_1&=2(r_1+r_2+r_3)+\frac{4(r_2-r_1)K(m)}{E(m)-K(m)},
\\
v_2&=2(r_1+r_2+r_3)-\frac{4(r_2-r_1)(1-m^2)K(m)}{E(m)-(1-m^2)K(m)},
\\
v_3&=2(r_1+r_2+r_3)+\frac{4(r_3-r_2)K(m)}{E(m)}
\end{aligned}
\end{equation}
and the cylindrical terms can be expressed in the form below, with the assumption $S_{i}=1-4r_{i},\quad (i=1,2,3)$ for simplicity in processes,
\begin{equation}\label{eqC23}
\begin{array}{lcl}
h_1=-\frac{S_1 E}{E-K}-\frac{\sqrt{S_1}(\sqrt{S_2}+\sqrt{S_3})(\sqrt{S_1}+\sqrt{S_2}\sqrt{S_3})\Pi}{(E-K)(S_1-S_3)}\\+\frac{K\sqrt{S_1}\Big(S_1+\sqrt{S_1}+S_2+\sqrt{S_2}+(\sqrt{S_2}-\sqrt{S_1})\sqrt{S_3}\Big)}{2(E-K)(\sqrt{S_1}-\sqrt{S_3})},\\
\\
h_2=-\frac{\sqrt{S_2}(\sqrt{S_2}+\sqrt{S_3})(\sqrt{S_2}+\sqrt{S_1}\sqrt{S_3})\Pi}{E(S_1-S_3)+K(S_3-S_2)}\\+\frac{E(S_3-S_1)S_2}{E(S_1-S_3)+K(S_3-S_2)}+K\sqrt{S_2}(\sqrt{S_2}+\sqrt{S_3})\\\times\frac{(S_1+S_2-\sqrt{S_2}(-1+\sqrt{S_3})+\sqrt{S_1}(1+\sqrt{S_3}))}{2(E(S_1-S_3)+K(S_3-S_2))},\\
\\
h_3=-S_3+\frac{\sqrt{S_3}(\sqrt{S_2}+\sqrt{S_3})(\sqrt{S_1}\sqrt{S_2}+\sqrt{S_3})\Pi}{E(S_3-S_1)}\\
+\frac{\sqrt{S_3}(\sqrt{S_2}+\sqrt{S_3})(1+\sqrt{S_1}+\sqrt{S_2}-\sqrt{S_3})K}{2(\sqrt{S_1}-\sqrt{S_3})E}.
\end{array}
\end{equation}

In Eq. (\ref{eqC22}), $v_{i}$'s are the Whitham characteristic velocities for the defocusing Gardner equation \cite{Kam12}. Also, $K=K(m)$, $E=E(m)$ and $\Pi=\Pi(n,m)$ denote the complete elliptic integrals of the first, second and the third kinds, respectively. Properties of these complete elliptic integrals are listed in the Appendix. \par
Note that Eq. \eqref{eqC21} reduces to a diagonal system in the absence of cylindrical terms, i.e. $t_0\rightarrow \infty$, that agrees with the Whitham system for the defocusing Gardner equation \cite{Kam12}.
\par
The expressions of the module, $m$, the wave number $k$ and the phase speed $V$ in terms of Riemann variables are as follows:
\begin{equation}\label{eqC24}
\begin{aligned}
m=\frac{\sqrt{r_2-r_1}}{\sqrt{r_3-r_1}},\hspace{0.5cm}k=\frac{\sqrt{r_3-r_1}}{2K(m)},
\\
V=2(r_1+r_2+r_3).
\end{aligned}
\end{equation}

%******************************************************
\section{\label{sec:level4} Numerical solution of the cylindrical gardner equation}
In this section, the direct numerical solution associated with the cG equation is examined. Then we will compare this numerical solution with the corresponding Whitham modulation system in the next section.\par
The defocusing Gardner equation,
\begin{equation}\label{eqD1}
f_{t}+6ff_{\eta}-6f^2f_{\eta}+\epsilon^2f_{\eta\eta\eta}=0
\end{equation}
with the initial condition \eqref{eqB12} depending  on the positions of the initial step parameters $f^-$ and  $f^+$ is considered in \cite{Kam12}. According to the analysis, wave structures with different cases such as dispersive shock wave (undular bore), rarefaction wave, solibore, reversed rarefaction wave have been observed depending on the choice of the initial step parameters $f^-$ and $f^+$. However, because of our study is related to dispersive shock waves, other cases are outside the scope of this paper. For the formation of a dispersive shock wave,  $f^+$ and $f^-$ must satisfy the inequality
\begin{equation}\label{eqD3}
f^+<f^-\leq1/2.
\end{equation}
\par
In our numerical simulations, we use a numerical procedure which is useful for problems with fixed boundary conditions. However, for the cG equation,
\begin{equation}
\label{eqD4}
f_{t}+6f f_{\eta}-6f^{2}f_{\eta}+\frac{1}{2t+t_0}f+\epsilon^2 f_{\eta\eta\eta}=0,
\end{equation}
$f$ satisfies the boundary conditions which are given for non-increasing type initial conditions as 
\begin{equation}
\label{eqD5}
   f\to R(t)f^- \text{ as } \eta\to-\infty \text{ and } f\to R(t)f^+ \text{ as } \eta\to\infty, 
\end{equation}
where $R(t)=\sqrt{\frac{t_0}{2t+t_0}}$ with $t_0=\frac{1}{\tilde{c}}$. Since these boundary conditions are functions of  $t$, we use the transformation
\begin{equation}
\label{eqD6}
 f=R(t)\psi
\end{equation}
and transform  Eq. (\ref{eqD4}) into the following Gardner equation with variable coefficients;
\begin{equation}
\label{eqD7}
 \psi_t+6R(t)\psi\psi_\eta-6(R(t))^2\psi^2\psi_\eta+\epsilon^2\psi_{\eta\eta\eta}=0.
\end{equation}
This equation has the constant left boundary condition with $\psi_-=f^-$ and the right boundary condition is $\psi_+=f^+$.
We use a modified version of the exponential time differencing fourth order Runge-Kutta (ETDRK4) method \cite{Trefethen}. For the required spectral accuracy of the method, the initial condition must be smooth and periodic. However, the step initial condition (\ref{eqB2}) for $u$ or equivalently for $f$  is non-periodic. To deal with this problem, we differentiate the equation (\ref{eqD7}) with respect to $\eta$ and define $\psi_{\eta}=z$. We obtain
\begin{equation}
\label{eqD8}
z_{t}+6R(t)(\psi z)_{\eta}-6(R(t))^2(\psi^2 z)_{\eta}+\epsilon^2 z_{\eta\eta\eta}=0. 
\end{equation}
The initial condition is regularized with the analytic function \cite{Ablowitz09}
\begin{equation}
\label{eqD9}
z(\eta,0)=-\frac{\hat{C}}{2}sech^2(\hat{C}\eta), 
\end{equation}
where $\hat{C}>0$ is large parameter. Therefore, the initial condition is now periodic and smooth, which is convenient for the numerical method.
 \par
To work on the Fourier space, we rewrite Eq. (\ref{eqD8}) as
\begin{equation}
\label{eqD10}
z_{t}=\boldsymbol{L}\hat{z}+6R(t)\boldsymbol{N_1}(\hat{z},t)-6(R(t))^2\boldsymbol{N_2}(\hat{z},t),
\end{equation}
where, $\hat{z}=\mathcal{F}(z)$ is the Fourier transform of $z$, $\boldsymbol{L}$ is the linear term and $\boldsymbol{N_1}$, $\boldsymbol{N_2}$ are the nonlinear terms. The expressions of linear and nonlinear terms are given as
\begin{equation}
\label{eqD11}\hspace{-0.1cm}
\begin{aligned}
\boldsymbol{L}\hat{z}&\!=\!-i\epsilon^2k^3\hat{z},
\\
\boldsymbol{N_1}(\hat{z},t)&\!=\!-ik\mathcal{F}\Big(\big(\int_{-L}^{\eta}\mathcal{F}^{-1}(\hat{z})d\eta'\!+\!\psi_-\big)
\mathcal{F}^{-1}(\hat{z})\Big),
\\
\boldsymbol{N_2}(\hat{z},t)&\!=\!-ik\mathcal{F}\Big(\big(\int_{-L}^{\eta}\mathcal{F}^{-1}(\hat{z})d\eta'\!+\!\psi_-\big)^2
\mathcal{F}^{-1}(\hat{z})\Big).
	\end{aligned}
\end{equation}
Consequently, equation (\ref{eqD8}) is solved numerically via Eqs (\ref{eqD11}) on a finite spatial domain $[-L,L]$. For the method ETDRK4, we take the number of Fourier modes in space as $N=2^{12}$, the domain size is $L=40$ and the time step is $h=10^{-3}$. Furthermore, the parameters $\tilde{c}^{-1}=t_0=10$,\quad $\epsilon^2=10^{-3}$ and $\hat{C}=10$.\par
The numerical solutions of the Gardner equation and the cG equation at $t=10$ are presented in Fig.1 and Fig.2, where the step parameters are chosen as $f^-=0.4$ and $f^+=0.1$.
\par 
In the next section, DSW solutions obtained by the numerical solutions of the modulation equations  and the direct numerical simulations of the Gardner and the cG equation will be compared. This allows us to understand the underlying structure of the DSWs in the cG equation.

\begin{figure}[H]
\includegraphics[width=\linewidth]{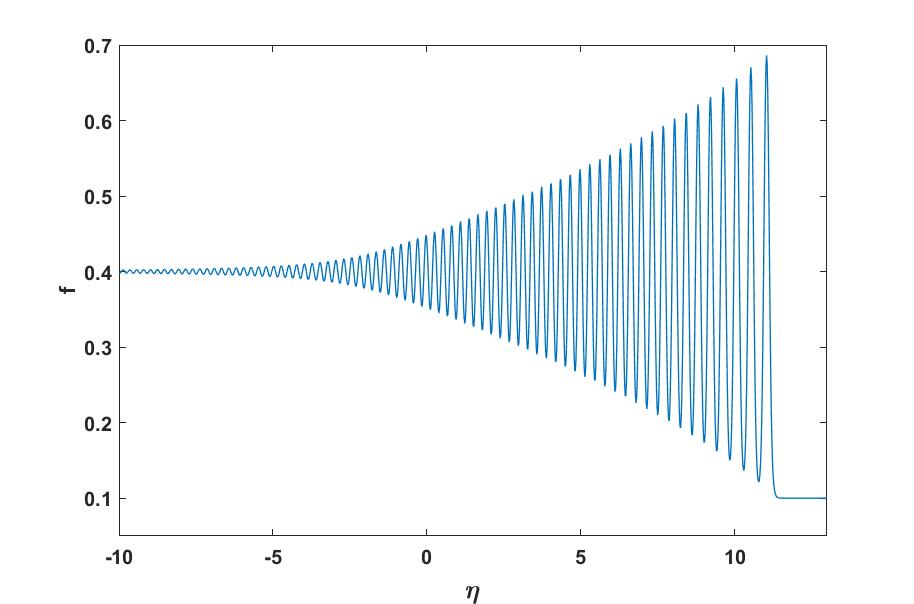}
  \caption{Numerical solution of Gardner quation at t=10 with the initial condition \eqref{eqB12} where $f^-=0.4$ and $f^+=0.1$. Here $t_0=10$ and $\epsilon^2=0.01$. }
 \label{fig:G_num}
\end{figure}

\begin{figure}[H]
\includegraphics[width=\linewidth]{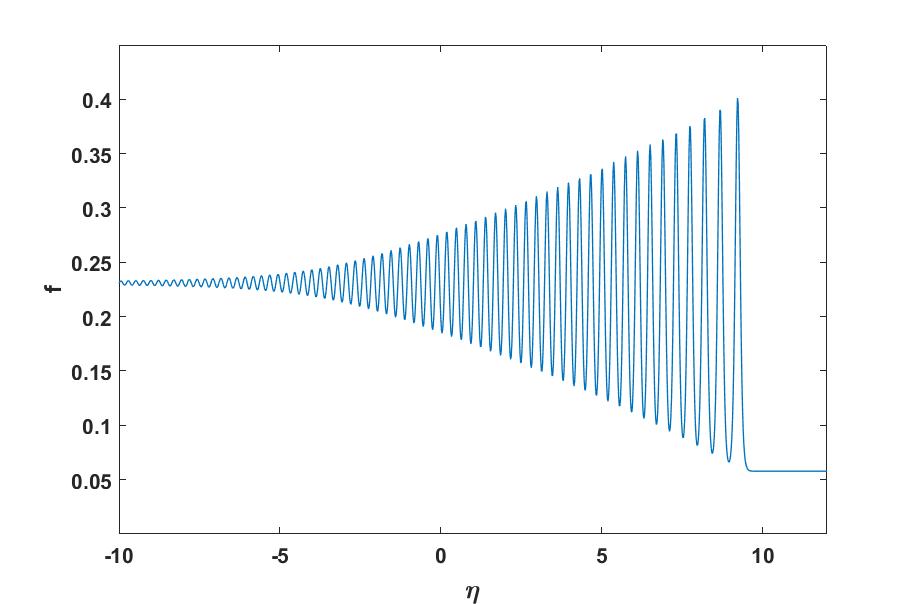}
  \caption{Numerical solution of cG quation at t=10 with the initial condition \eqref{eqB12} where $f^-=0.4$ and $f^+=0.1$. Here $t_0=10$ and $\epsilon^2=0.01$.}
 \label{fig:cG_num}
\end{figure}

%*******************************************************
\section{\label{sec:level5} Dispersive shock waves in the cylindrical gardner equation and comparison with numerical results}
In this section we obtain the  DSW solution of the cG equation by using the solution of the Whitham modulation system given in (\ref{eqC21}). Then we compare this DSW solution with direct numerical simulation of the cG equation. Since the Whitham system \eqref{eqC21} is not in diagonal form,  it is difficult to get its analytical solution. Therefore we solve the corresponding Whitham system by using a numerical method.
\par
 It is  clear that without the cylindrical term, the system in (\ref{eqC21})  reduces to the Whitham system for the Gardner equation \cite{Kam12}. To observe the effect of the cylindrical term on the DSW solution, the numerical solution of the corresponding modulation equations for the Gardner equation will be also investigated. For this purpose, we use  a first order hyperbolic PDE solver based on $\textrm{MATLAB}^®$ by Shampine \cite{Shampine} and choose a two-step variant of the Lax-Wendroff method with a nonlinear filter \cite{Nonfilter}. 
 \begin{figure}[H]
\includegraphics[width=\linewidth]{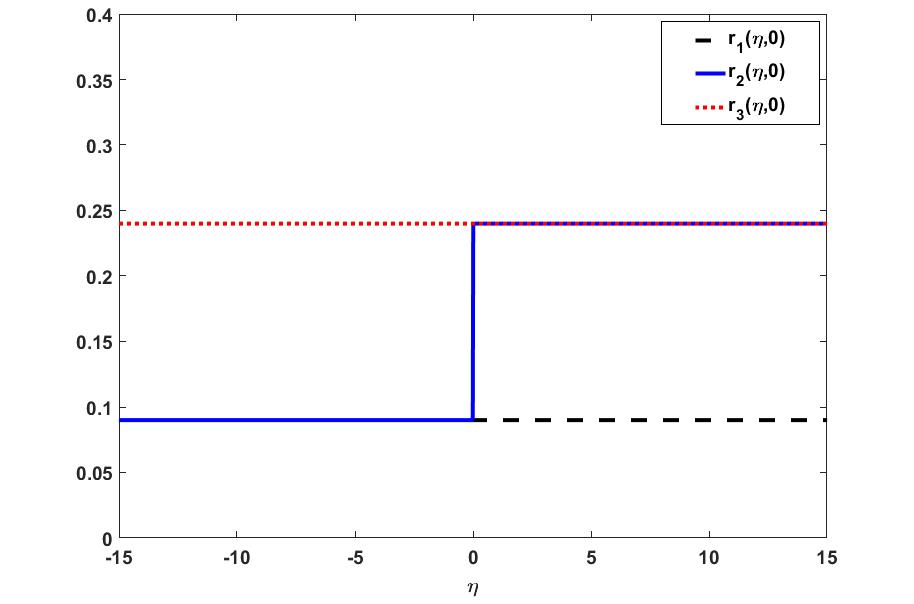}
\caption{Initial values of Riemann variables.}
  \label{fig:riemann_baslangic}
\end{figure}
\par
In order to solve Whitham modulation equations numerically, we must first obtain boundary conditions for the Whitham system. The boundary conditions for the Gardner equation and associated Whitham system of this equation remain constant at both ends of the domain, since they do not contain time dependency (see Fig. 1). However, the boundary conditions for the cG equation and associated Whitham system are the functions of time (see Fig. 2). The boundary conditions for the cG equation can be seen in Eq. (\ref{eqD5}). In order to find the boundary conditions for the Whitham system (\ref{eqC21}), we solve the ODE system obtained from Eq. (\ref{eqC21}) analytically by neglecting the spatial variable. The reduced ODE system is solved with the initial conditions (\ref{eqE1}) at both ends separately. The initial values of Riemann variables for the Whitham system is given as follows (see Fig. 3):
\begin{equation} \label{eqE1}
\begin{aligned}
 r_1(\eta,0)&=0.09,\quad  r_3(\eta,0)=0.24
 \\
 r_2(\eta,0)& = \begin{cases}
    0.09 &  \eta\leq0; \\
    0.24 &  \eta>0,  
  \end{cases}
  \end{aligned}
\end{equation}

Since the structures of the analytic solutions are so complicated, the exact forms of these solutions of the reduced ODE system are not given here. The numerical solutions of Whitham systems including boundary conditions at $t=10$ are given in Fig. 4 for Gardner and cG equations.
\par In the numerical solutions of the Whitham systems, we use $N=2^{12}$ points for the spatial domain $[-40,40]$. Furthermore, the parameter $\tilde{c}^{-1}=t_0=10$ in the cG equation. 
\par
From Fig. 4, it should be noted that a difference between both the positions of the intersection points of the Riemann variables $r_1$ and $r_2$ at the trailing edges is observed when the behavior of Riemann variables of Gardner and cG equations are compared. Similar difference is observed at the positions of the intersection points of the Riemann variables 
$r_1$ and $r_2$ at the leading edge, too. These differences imply some results about the edge dynamics of DSWs in Gardner and cG equations. Both the leading edge and trailing edge of the DSW in the cG equation move slower than both the leading and trailing edges of the DSW in the Gardner equation. All these results agree with Fig. 5, Fig. 6 and an animation \cite{Animation}. In the animation \cite{Animation}, the propagation of DSWs in both Gardner and cG equations is between $t=0$ and $t=10$.
\par
Now, in order to compare the asymptotic (modulation theory) solution of the cG equation with the direct numerical solution, we first write the asymptotic solution, $f_0$, in terms of Riemann variables $r_i$ 
\begin{equation}
\label{eqE3}
\begin{split}
&f_0(\theta,\eta,t)=\frac{1}{2}(1-\sqrt{S_1}+\sqrt{S_2}-\sqrt{S_3})\\
&+\frac{(\sqrt{S_1}-\sqrt{S_2})(\sqrt{S_1}+\sqrt{S_3})cn^2(2(\theta-\theta_0)K,m)}{(\sqrt{S_1}+\sqrt{S_3})+(\sqrt{S_2}-\sqrt{S_1})sn^2(2(\theta-\theta_0)K,m)},
\end{split}
\end{equation}
where $S_{i}=1-4r_{i},\quad (i=1,2,3)$. By integrating (\ref{eqC3}), the rapid phase $\theta$ is obtained as:
\begin{equation}
\label{eqE5}
	\theta(\eta,t)=\int_{-L}^{\eta}\frac{k(\eta',t)}{\epsilon}d\eta'-\int_{0}^{t}\frac{k(\eta,t')V(\eta,t')}{\epsilon}dt'.
\end{equation}
 The asymptotic solution of the cG equation, $f_0$, is obtained by using (\ref{eqE3}) and the formula (\ref{eqE5}) of $\theta$.
\begin{figure}[H]
 %\begin{subfigure}{\linewidth}
\includegraphics[width=\linewidth]{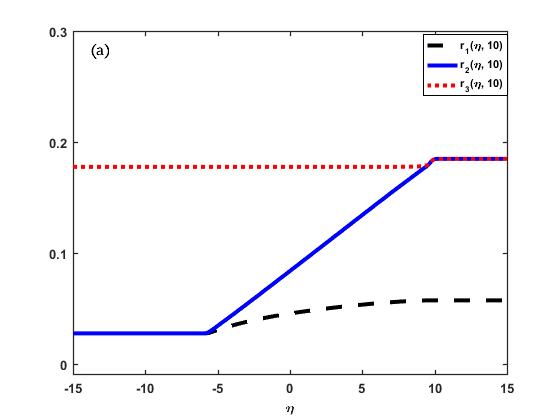}
 %\end{subfigure}
  %\begin{subfigure}{\linewidth}
    \includegraphics[width=\linewidth]{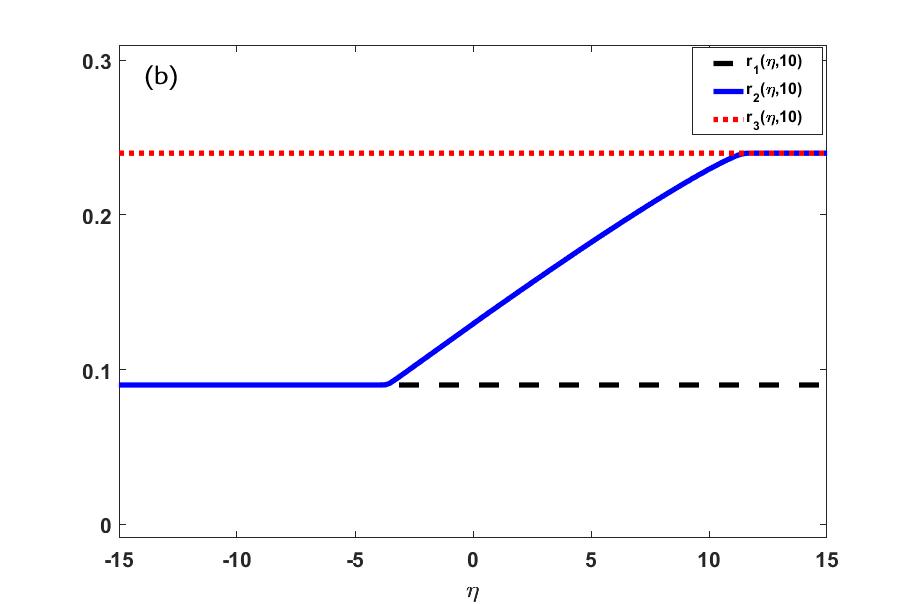}
  %\end{subfigure}
  \caption{(a) Riemann variables at t=10 which are found by numerical solutions of \eqref{eqC21} (a) for cG eq.\quad(b) for Gardner eq. Here we take $t_0=10$.}
  \label{fig:riemann}
\end{figure} 

Thus, the DSW solutions can be generated at any time for both Gardner and cG from the Riemann variables $r_i$'s using Eqs. \eqref{eqE3} and \eqref{eqE5}. The direct numerical solution of the cG equation is plotted and compared with the asymptotic solution in Fig.5. Accordingly, it can be observed from Fig.5 that, the leading edge amplitude and wavelength of oscillations are compatible in both of asymptotic and numerical solutions. Thus, Whitham modulation theory enables us to obtain correct and appropriate approaches for dispersive shock waves in the cG equation.

In the numerical approach, the phase shift $\theta_0$ has been arbitrarily chosen as it is compatible with direct numerical simulations. For this adjustment, we find the mean value of the DSW, that is, we compute the average of the leading hump (the largest amplitude soliton) and the trailing edge. Then $\theta_0$ is selected as the center of the nearest wave determined from the asymptotic solution which is identical to the corresponding hump in the direct numerical simulations. For the cG equation, the average is approximately $(0.4001+0.2309)/2=0.3155$ and the asymptotic solution has a hump in the middle (center) region with a value of amplitude $0.2958$.
\begin{figure}[H]
\includegraphics[width=\linewidth]{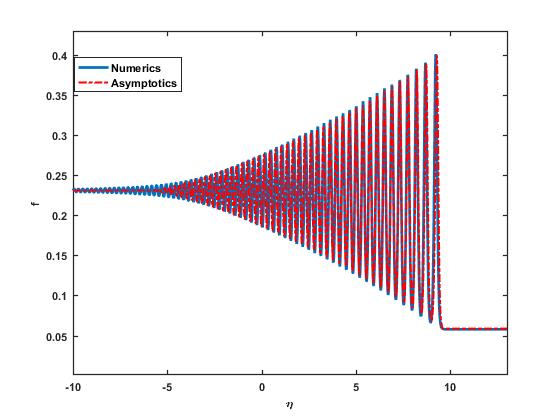}
  \caption{Numerical and asymptotic solutions of cG equation at t=10 with the initial condition \eqref{eqB12} where $f^-=0.4$ and $f^+=0.1$. Here $t_0=10$ and $\epsilon^2=0.01$.}
 \label{fig:cakisik}
\end{figure}
\par
The similar analysis can be done for the Whitham system of Gardner equation to make a comparison. By using the same procedure for this equation, the asymptotic solution can be obtained by taking the cylindrical term $\tilde{c}=0$. The initial values of the Riemann variables are the same as cG equation (Fig. 3). Then, the direct numerical simulation of Gardner equation and the asymptotic solution at $t=10$ are compared in Fig. 6.
\begin{figure}[H]
\includegraphics[width=\linewidth]{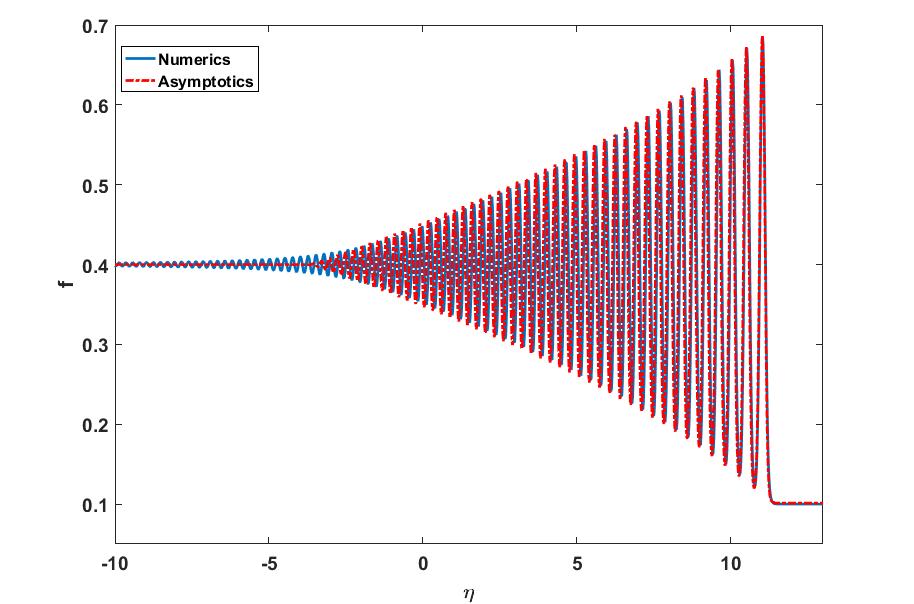}
  \caption{Numerical and asymptotic solutions of Gardner eq. at t=10 with the initial condition \eqref{eqB12} where $f^-=0.4$ and $f^+=0.1$. Here $t_0=10$ and $\epsilon^2=0.01$.}
  \label{fig:cg_cakisik.eps}
\end{figure} 
For the Gardner equation, the average of the leading hump and the trailing edge is approximately $(0.6878+0.4)/2=0.5439$ and a hump in the middle region has an amplitude value $0.5199$. As we can see, the direct numerical simulations and asymptotic solutions are compatible for both Gardner and cG equations. 
\par  
However, one needs to proceed to higher order terms in the asymptotic expansion (\ref{eqC2}) to achieve better results for the phase term $\theta_0$. But, this is outside the scope of this paper.         

\section{\label{sec:level6}Conclusions}
We investigate the DSW solution of the cG equation by using Whitham modulation theory. cG equation is derived from the reduction of the (2+1) dimensional Gardner-KP equation with step-like initial condition along a parabolic front. The formation of the DSW depends on the selection of the initial step parameters $f^{-}, f^{+}$. We require that  $0\leq f^+<f^-\leq1/2$ in order to generation the DSW solution in cG equation. Then by using the method of multiple-scale we obtain a system of quasilinear modulation equations describing slow evolution of parameters in the periodic solution of a dispersive nonlinear partial differential equation. We solve the corresponding Whitham modulation equations numerically and compare these results with direct numerical solutions of the cG equation. A good agreement is found between these numerics except the negligible phase term. Effect of this phase term can be analysed by considering higher order terms; but this is outside the scope of this study. In order to observe the contribution of the cylindrical term in cG equation, a DSW formation of the Gardner equation is also considered in this study. The conclusion we get from the simulations is that the amplitude decreases over time for the cylindrical equation in the trailing and leading edge. Observations can be made similarly for other t values.
\par
In \cite{Kam12}, other wave type solutions of the Gardner equation such as rarefaction wave, solibore, reversed rarefaction wave was studied depending on the choice of the initial step parameters $f^-$ and $f^+$. Similar analysis can be performed for the cG equation. We address this investigation for near future studies.
\par
The method introduced in this study for the reduction of the Gardner-KP works only for a special choice of initial front, e.g. a parabolic front. However, Ablowitz et. al. generalized the Whitham Theory to find  DSW solutions of the KP- type equations with a general class of initial conditions \cite{Ablowitz18}. In this study, a (2+1) dimensional Whitham modulation system was derived, which describes the slow modulations of the periodic solutions of the corresponding KP- type equations. To our knowledge any quantitative results for the solutions of derived (2+1) dimensional Whitham systems have not been reported yet. The Gardner- KP equation belongs to the KP class which was investigated in \cite{Ablowitz18}. Our results about the cG equation can be used as a test subject to verify the DSW solutions obtained from the solution of the (2+1) dimensional Gardner-KP Whitham system.

\section*{ACKNOWLEDGMENTS}
This research was supported by the Istanbul Technical University Office of Scientific Research Projects (ITU BAPSIS), under grant TGA-2018-41318. We thank D.E. Baldwin for MATLAB codes of version of the ETDRK4 method that we use in the study.

\setcounter{section}{0}
\setcounter{equation}{0}
\renewcommand{\theequation}{\thesection\arabic{equation}}
\appendix
\section*{Appendix: Complete elliptic integrals}
\renewcommand{\thesection}{A}

In this Appendix, some properties of complete elliptic integrals used in the study will be listed.\par
The first kind complete elliptic integral have the expansion: 
\begin{equation}
\label{eqA1}
\begin{split}
K(m)=&\frac{\pi}{2}\Big(1+\frac{m}{4}+\frac{9}{64}m^2+\cdots\\
&+\Big(\frac{1\cdot3\cdot\cdot\cdot(2n-1)}{2\cdot4\cdots2n}\Big)^2m^n+\cdots\Big)
\end{split}
\end{equation}
Series expansion of the second kind elliptic integral:  
\begin{equation}
\label{eqA2}
\begin{split}
E(m)=&\frac{\pi}{2}\Big(1-\frac{m}{4}-\frac{3}{64}m^2-\cdots\\
&-\frac{1}{2n-1}\Big(\frac{1\cdot3\cdots(2n-1)}{2\cdot4\cdots2n}\Big)^2m^n-\cdots\Big), 
\end{split}
\end{equation}
for $|m|<1$. \par  
The compete elliptic integral of the third kind has the following behavior
\begin{equation}
\label{eqA3}
\Pi(n,m)=\frac{\pi}{2} \quad \textrm{when}\quad n=0,\quad m=0
\end{equation}

\begin{equation}
\label{eqA4}
\frac{\Pi(n,m)}{K(m)}\approx\frac{1}{1-n} \quad \textrm{when}\quad m \quad\textrm{is close to 1.}
\end{equation}
The following are the derivative formulas:
\begin{equation}
\label{eqA5}
\frac{dK(m)}{dm}=\frac{E(m)-(1-m) K(m)}{2 (1-m) m},
\end{equation}
\begin{equation}
\label{eqA6}
\frac{dE(m)}{dm}=\frac{E(m)-K(m)}{2m},
\end{equation}
\begin{equation}
\label{eqA7}
\frac{d\Pi(n,m)}{dm}=\frac{E(m)-(1-m)\Pi(n,m)}{2(1-m)(m-n)},
\end{equation}
\begin{equation}
\label{eqA8}
\frac{d\Pi(n,m)}{dn}=\frac{nE(m)+(m-n)K(m)+(n^2-m)\Pi(n,m)}{2n(1-n)(n-m)}.
\end{equation}
%\subsection{\label{sec:citeref}}

\end{document}